\documentclass[amssymb,twoside,12pt]{article}
\usepackage{latexsym,amsmath}
\usepackage{hyperref}
\usepackage{amsfonts}
\usepackage{color}
\usepackage{amsmath}
\usepackage{enumerate}
\usepackage{natbib}
\usepackage{graphics,graphicx}
\usepackage[center]{titlesec}
\begin{document}
\setcounter{page}{1}
\begin{center}

\vspace{0.4cm} {\large{\bf\bf Plane Symmetric Cosmological Model with Quark and Strange Quark Matter in f(R,T) Theory of Gravity
}} \\

\vspace{0.4cm}
P. K. Agrawal, D. D. Pawar \\
School of Mathematical Sciences,\\
Swami Ramanand Teerth Marathwada University, Nanded-431606, India\\
E-mail: agrawalpoonam299@gmail.com, dypawar@yahoo.com \\
\vspace {0.2cm}
\end{center}
\begin{abstract}
 We studied plane symmetric cosmological model in the presence of quark and strange quark matter with the help of f(R,T) theory. To decipher solutions of Plane symmetric space-time, we used power law relation between scale factor and deceleration parameter. We considered the special law of variation of Hubble's parameter proposed by Berman (Nuovo Cimento B 74, 182, 1983) which yields constant deceleration parameter. We also discussed the physical behavior of the solutions by using some physical parameters.
\end{abstract}
\def\baselinestretch{1.5}
\allowdisplaybreaks
\section{Introduction}
\paragraph*{}Modern astrophysical observations point out that present expansion of the universe is an accelerated epoch. The most fascinating evidence for this is found in measurements of type-Ia supernovae(SNela) \citep{1538-3881-116-3-1009,0004-637X-517-2-565}, which is supported by eminent observations like \citep{PhysRevD.69.103501}. To explain this behavior of the universe mysterious component of energy, now known as dark energy, plays a major role. However, the mechanism behind the accelerated expansion is still point of discussion. Modified theories of gravity have been evolved to explain such issue of current cosmic acceleration. f(R) \citep{PhysRevD.75.084031}, f(T) \citep{PhysRevD.79.124019} where T is the torsion scalar in teleparallel and f(R, T) \citep{PhysRevD.84.024020} where R is the Ricci scalar and T is the trace of the energy momentum tensor, are few types of modified theories of gravity.
\paragraph*{}f(R,T) gravity is proposed by \cite{PhysRevD.84.024020} which is the extension of standard general relativity where the gravitational Lagrangian is given by an arbitrary function of the Ricci scalar R and of the trace of the stress energy tensor T. \cite{Sharif2013} have obtained the thermodynamics at the apparent horizon of the FRW universe in f(R,T) theory.  \cite{Chandel2013} and \cite{Reddy2013} have studied spatially homogeneous and anisotropic Bianchi type-III space-time and LRS Bianchi type-II space-time respectively, with perfect fluid in f(R,T) theory of gravity. \cite{}  \cite{Pawar2016} and \cite{Pawar2015} have studied Bianchi type-I  cosmological model in the framework of f(R,T) theory of gravity. \cite{Pawar2016a} have discussed axially symmetric space-time in f(R,T) theory of gravity. \cite{Pawar2015a} have examined Bianchi type-I cosmological model in the presence of dark energy in f(R,T) theory of gravity.
\paragraph*{} During initial studies the universe was considered to be spherically symmetric and the matter distribution in it is isotropic and homogeneous. But in the recent studies of evolution, the picture has changed. The assumptions of spherically symmetric and the isotropy do not hold strictly near the big bang singularity. From this point of view to study early days inhomogeneities many authors considered plane symmetry. Inhomogeneous plane-symmetric model was first studied by \cite{Taub1951,Taub1956} and later by \cite{Tomimura}, \cite{Szekeres}, \cite{Pradhan}, \cite{Singh}, \cite{Bayaskar}, \cite{Taruya}, \cite{Pawar2008, Pawar2009, Pawar2009b}, etc. In this work we study the quark and the strange quark matter in the f(R, T) theory for plane symmetric space-time.
\paragraph*{}A lot of work has been done in f(R,T) theory of gravity using various kinds of matter. Two such kinds are Quark and Strange quark. The fundamental particles of quark matter are bound together by the strong interaction. Quark matter is considered to exist at the center of neutron stars \citep{PhysRevLett.105.141101}, in strange stars \citep{0004-637X-572-2-996}, or even as small pieces of strange matter \citep{Weber2005193}. The quark matter is thought to be originated when the universe underwent a quark gluon phase transition for a few microseconds after the big bang \citep{ref1}. In 1984, \cite{PhysRevD.30.272} demonstrated that at a critical temperature $ T_{c} \equiv 100-200 MeV $ such transition could have led to the formation of quark nuggets made up of u, d and s quarks at larger density than normal nuclear matter density. Strange quark matter is developed with an equation of state (EoS) $p = \frac{(\rho - 4B_{c})}{3}$ based on the phenomenological bag model of quark matter in which quark constraint is described by an energy term proportional to the volume. In this equation $B_{c}$ is known as bag constant, which is difference between the energy density of the perturbative and non-perturbative quantum chromodynamics (QCD) vacuum. p and $\rho$ are thermodynamic pressure and energy density of the quark matter respectively. In this model quarks are thought as degenerate Fermi gas, which exist only in a region of space endowed with a vacuum energy density $B_{c}$ (called as the bag constant). In the frame of reference of this model, the quark matter is  possessed of mass-less u and d quarks, massive S quarks and electrons. In the simplified version of the bag model, it is assumed that quarks are mass-less and non-interacting. Hence, we have quark pressure $ p_{q} = \frac{\rho_{q}}{3}$, where $\rho_{q}$ is the quark energy density. The total energy density is $\rho = \rho_{q} + B_{c}$ and the total pressure is $p = p_{q} - B_{c}$.
\paragraph*{}Nowadays, the study of Quark and Strange quark matter is an interesting topic of research. Quark matter is studied in general relativity using various assumptions. \cite{doi:10.1142/S0218271804004451} have studied charged strange quark matter with the help of spherically symmetric space-time using conformal motion. \cite{Katore} have acquired FRW cosmological model with strange quark matter attached to the string cloud in general relativity. \cite{Yılmaz2012} have examined quark and strange quark matter in f(R) theory of gravity for Bianchi type I and V space-time. \cite{Khadekar} have obtained higher dimensional cosmological model in the presence of quark and strange quark matter. \cite{Adhav2015} have discussed the Kantowski-Sachs Cosmological model with quark and strange quark matter in f(R) gravity. \cite{Rao2015} have studied Bianchi type-VI$_{0}$ space-time with strange quark matter attached to string cloud. \cite{Sahoo, Santhikumar2015} have obtained axially symmetric cosmological model with string cloud universe containing strange quark matter. So by considering the importance of study of plane symmetry and Quark matter, we tried to consider plane symmetric cosmological model in both Quark and Strange quark matter. 
\paragraph*{}Motivated by above work, we studied Plane Symmetric Cosmological model in the presence of quark and strange quark matter in the frame work of f(R,T) theory of gravity. To obtain exact solutions of the Einstein's field equations we considered power law relation between scale factor and deceleration parameter. We also assumed the special law of variation of Hubble's parameter, which yields constant deceleration parameter. The paper is organized as follows. In Section 2 we described the necessary field equations of f(R,T) gravity. Section 3 contains metric and field equations. In section 4 we derived the solutions of the field equations. Section 5 describes the physical properties of the model. A brief discussion of the results and the conclusion is provided in section 6.
\section{Gravitational field equations of f(R,T) gravity}
The f(R,T) theory of gravity is one of the modified theories of gravity which is developed by \citep{PhysRevD.84.024020}. He has obtained the field equations of f(R,T) gravity from the Hilbert-Einstein principle. In this section we will discuss the brief review of this theory as follows:
\\The action for the modified f(R,T) gravity is given as
\begin{equation}\label{eq:x1}
  S = \frac{1}{16 \pi} \int f(R,T)\sqrt{-g}    d^4x + \int L_m\sqrt{-g}    d^4x
\end{equation}
  where f(R,T) is an arbitrary function of R and T, R is the Ricci scalar and T is the trace of energy momentum tensor of the matter $T_{ij}$. $L_{m}$ is the matter Lagrangian density. The energy momentum tensor $T_{ij}$ is defined as
 \begin{equation}\label{eq:x2}
 T_{ij}= \frac{-2\partial(\sqrt{-g}L_m)}{\sqrt{-g}\partial g^{ij}}
\end{equation}
  The trace is defined by $T = g^{ij}T_{ij}$.
  Here assume that the matter Lagrangian $L_{m}$ depends only on the metric tensor component $g_{ij}$ rather than its derivatives. In this case
 \begin{equation}\label{eq:x3}
 T_{ij} = g_{ij}L_{m} - 2\frac{\partial L_{m}}{\partial g^{ij}}
 \end{equation}
  The field equations of f(R,T) gravity are derived by varying the action S with respect to metric tensor $g_{ij}$,
  \begin{equation}\label{eq:x4}
  f_R(R,T)R_{ij} -\frac{1}{2}f(R,T)g_{ij} + (g_{ij}\square - \bigtriangledown_{i}\bigtriangledown_{j})f_R(R,T) =  8\pi T_{ij} - f_T(R,T)T_{ij} - f_T(R,T)\theta _{ij}
  \end{equation}
 Here

 \begin{equation}\label{eq:x5}
 \theta_{ij} = -2T_{ij} + g_{ij}L_{m} -2g^{\alpha \beta} \frac{\partial^{2}L_{m}}{\partial g^{ij}\partial g^{\alpha \beta}}
 \end{equation}
 where
  $f_R(R,T)= \frac{\partial f(R,T)}{\partial R}, f_T(R,T) = \frac{\partial f(R,T)}{\partial T}, \square =\bigtriangledown ^{\mu}\bigtriangledown_{\mu}$

  where $\bigtriangledown_{\mu}$ is the co-variant derivative.\\
On contracting eqn~\eqref{eq:x4} , it gives a relation between Ricci scalar R and the trace T of energy momentum tensor.
\begin{equation}\label{eq:x6}
f_R(R,T)R + 3\square f_R(R,T) - 2f(R,T)= 8 \pi T - f_T(R,T)(T+ \theta),
\end{equation}
where $\theta = \theta^{i}_{i}$.\\
The stress energy tensor of the matter is given by
\begin{equation}\label{eq:x7}
T_{ij} = (\rho + p)u_{i}u_{j} - pg_{ij},
\end{equation}
where $\rho $ and p are energy density and pressure of the fluid respectively. The matter Lagrangian can be taken as $L_{m}= -p $. $ u^{i} = (0,0,0,1)$ is the four-velocity in co-moving co-ordinates and satisfies the condition $u^{i}u_{i} = 1$ and $ u^{i}   \bigtriangledown_{j} u_{i} = 0$. Using $L_{m} = -p$ in eqn~\eqref{eq:x5}, we get the variation of the stress energy of a perfect fluid as follows
\begin{equation}\label{eq:x8}
\theta_{ij} = -2T_{ij} -p g_{ij}.
\end{equation}

Due to the physical nature of the matter field, field equations depend on the tensor $\theta_{ij}$. As f(R,T) gravity depends on the matter field, we get different theoretical models corresponding to different matter contributions for f(R,T) gravity. The three classes of these models are given as follows
\begin{equation}\label{eq:x9}
f(R,T)= \begin{cases}  R + 2f(T) \\ f_{1}(R) +f_{2}(T) \\ f_{1}(R) +f_{2}(R)f_{3}(T)\end{cases}
\end{equation} 
We considered first class, i. e. $f(R,T) = R + 2f(T)$, where $f(T)$ is an arbitrary function of stress energy tensor of matter and given by $f(T) = \lambda T, $ where $ \lambda $ is a constant. eqn~\eqref{eq:x4} gives the gravitational field equations of $f(R,T)$ gravity as follows.
\begin{equation}\label{eq:x10}
  G_{ij}= R_{ij} - \frac{1}{2}Rg_{ij}
  =  8 \pi T_{ij} - 2f'(T)T_{ij} - 2f'(T)\theta_{ij} + f(T)g_{ij},
  \end{equation}
where prime denotes differentiation with respect to the argument. If matter source as a perfect fluid then the field equations becomes
\begin{equation}\label{eq:x11}
  G_{ij} = R_{ij} - \frac{1}{2}Rg_{ij}
  =  8 \pi T_{ij} +2f'(T)T_{ij} + [2pf'(T) + f(T)]g_{ij}
  \end{equation}
\section{Metric and the field equations}
The plane symmetric space-time is considered in the form
\begin{equation}\label{eq:x12}
ds^{2} = dt^{2} - A^{2}(dx^{2} + dy^{2})- B^{2}dz^{2}
\end{equation}
where A and B are cosmic scale factors and are functions of t.\\
The corresponding Ricci scalar is given by
\begin{equation}\label{eq:x13}
R = \frac{2\dot{A}^{2}}{A^{2}} + 4 \frac{\ddot{A}}{A}+ 4\frac{\dot{A}\dot{B}}{AB}+  2\frac{\ddot{B}}{B}
\end{equation}
where overhead dot (.) denotes derivative with respect to time 't'.
The spatial volume V of the universe is defined as
\begin{equation}\label{eq:x14}
V = A^{2}B
\end{equation}
The generalized mean Hubble's parameter H is given as
\begin{equation}\label{eq:x15}
H=\frac{1}{3}\frac{\dot{V}}{V}= \frac{\dot{a}}{a}=\frac{1}{3}(H_{1}+ H_{2} + H_{3})
\end{equation}
where $H_{1} = H_{2} = \frac{\dot{A}}{A}, \hspace{.2 cm} H_{3} = \frac{\dot{B}}{B}$ are the directional Hubble parameter in the direction of x, y, z axes respectively.\\
The expansion scalar $\theta$ is given by
\begin{equation}\label{eq:x16}
\theta = 3H = \frac{2\dot{A}}{A} + \frac{\dot{B}}{B}
\end{equation}
Shear scalar $\sigma$ and the mean anisotropic parameter $A_{m}$ is defined as
\begin{equation}\label{eq:x17}
\sigma^{2} = \frac{1}{2}[\sum_{i=1}^{3} H_{i}^{2} -3 H^{2}] = \frac{3}{2}A_{m}H^{2}
\end{equation}
and
\begin{equation}\label{eq:x18}
A_{m} = \frac{1}{3}\sum_{i=1}^{3} \left(\frac{\triangle H{i}}{H}\right)^{2}
\end{equation}
where $\triangle H_{i} = H_{i} - H.$
\\The energy momentum tensor for quark matter is taken as
\begin{equation*}
T_{ij}^{(Quark)} = (\rho + p )u_{i}u_{j} - pg_{ij}
\end{equation*}
or
\begin{equation}\label{eq:x19}
T_{ij}^{(Quark)} = diag(\rho, -p, -p, -p ),
\end{equation}
where $\rho = \rho_{q} + B_{c}$ is the quark matter total energy density and $p = p_{q} - B_{c}$ is the quark matter total pressure and $u_{i}$ is the four-velocity vector such that $u_{i}u^{i} = 1$.\\
EoS parameter for quark matter is defined as
\begin{equation}\label{eq:x20}
p_{q}= \epsilon \rho_{q}, \hspace{.5 cm} 0 \le \epsilon \le 1
\end{equation}
For strange quark matter, the linear equation of state is given by
\begin{equation}\label{eq:x21}
p = \epsilon(\rho - \rho_{0})
\end{equation}
where $\epsilon$ is a constant and $\rho_{0}$ is the energy density at zero pressure.\\
 When $\epsilon = \frac{1}{3}$ and $\rho_{0} = 4B_{c}$ the above linear equation of state is reduced to the following EoS for strange quark matter in the bag model.
\begin{equation}\label{eq:x22}
p = \frac{\rho - 4B_{c}}{3}
\end{equation}
where $B_{c}$ is the bag constant.\\
In the co-moving co-ordinate system, the field eqn~\eqref{eq:x11} for metric \eqref{eq:x12} with the help of eqn~\eqref{eq:x19} can be written as
\begin{equation}\label{eq:x23}
\frac{ \ddot{A}}{A} + \frac{\ddot{B}}{B} + \frac{\dot{A}\dot{B}}{AB} = p_{q} -B_{c} - \lambda(\rho_{q} + 4B_{c} -3p_{q})
\end{equation}
\begin{equation}\label{eq:x24}
 \frac{ \dot{A}^{2}}{A^{2}} + 2\frac{\ddot{A}}{A} = p_{q} -B_{c} - \lambda(\rho_{q} + 4B_{c} -3p_{q})
\end{equation}
\begin{equation}\label{eq:x25}
 \frac{ \dot{A}^{2}}{A^{2}} +2 \frac{\dot{A}\dot{B}}{AB} = -\rho_{q} -B_{c} - \lambda(3\rho_{q} + 4B_{c} -p_{q})
\end{equation}
\section{Solution of the field equations}

Consider the special law of variation of Hubble's parameter proposed by \citep{Berman, B} which yields the constant deceleration parameter given by the relation
\begin{equation}\label{eq:x26}
q = \frac{-a\ddot{a}}{\dot{a}^{2}}
\end{equation}

In eqn~\eqref{eq:x26} `a' is the average scale factor. From eqn~\eqref{eq:x12} of given metric `a' is given by
\begin{equation}\label{eq:x27}
a^{3} = V \hspace{.5 cm} \Rightarrow  \hspace{.5 cm} a= (A^{2}B)^{\frac{1}{3}}
\end{equation}

The deceleration parameter can be constant if we relate the Hubble parameter H to the average scale factor `a', as considered by \citep{Berman} in solving FRW models, by the relation
\begin{equation}\label{y1}
H=la^{-m}=l(A^{2}B)^{\frac{-m}{3}}
\end{equation} 
where $l$ and $m$ are constants.
Using eq~\eqref{eq:x15}, we can re-write the above equation as
\begin{equation*}\label{y2}
\dot{a}=la^{-m+1}
\end{equation*}
\begin{equation*}\label{y3}
\ddot{a}=-l^2(m-1)a^{-2m+1}
\end{equation*}
Substituting the values of $\dot{a}$ and $\ddot{a}$ in eqn~\eqref{eq:x26} we get,
\begin{equation}
q=m-1
\end{equation}  

This equation produces a constant value for deceleration parameter and can have both positive as well as negative values. Positive value of deceleration parameter gives the standard deceleration model while the negative value results into inflation or the accelerating universe.\\

On solving eqn~\eqref{eq:x26} we get as
\begin{equation}\label{eq:x28}
a = (c_{1}t + c_{2})^{\frac{1}{q+1}}, \hspace{.5 cm} q\neq -1
\end{equation}
provided $c_{1} \neq 0$ and $c_{2}$ are constants of integration.\\
Equations \eqref{eq:x27} and eqn~\eqref{eq:x28} will give
\begin{equation}\label{eq:x29}
A^{2}B =(c_{1}t + c_{2})^{\frac{3}{q+1}}, \hspace{.5 cm} q\neq -1
\end{equation}
Now to solve system completely, we assume shear scalar $ \sigma $ is proportional to the expansion scalar $ \theta $ which gives a linear relationship between the directional Hubble parameters $H_{x}$ and $H_{z}$ as  $n H_{x} = H_{z}$. This assumption gives an anisotropic relation between the scale factors A and B as following
\begin{equation}\label{eq:x30}
B = A^{n}
\end{equation}
where $n\neq 1$ is an arbitrary constant. If $ n  =  1$,  the 
model  becomes  isotropic  model  otherwise it becomes anisotropic.\\
Hence eqn~\eqref{eq:x29} will imply
\begin{equation}\label{eq:x31}
A =(c_{1}t + c_{2})^{\frac{3}{(q+1)(n +2)}}, \hspace{.5 cm}  and \hspace{0.5 cm} B =(c_{1}t + c_{2})^{\frac{3n}{(q+1)(n+2)}}, \hspace{.5 cm} q\neq -1 \hspace{.2 cm}, \hspace{.2 cm} n\neq -2
\end{equation}
Thus the metric eqn~\eqref{eq:x12} with the help of eqn~\eqref{eq:x31} can be written as
\begin{equation}\label{eq:x32}
ds^{2} = dt^{2} - [c_{1}t + c_{2}]^{\frac{6}{(q+1)(n+2)}}(dx^{2} + dy^{2}) - [c_{1}t + c_{2}]^{\frac{6n}{(q+1)(n+2)}}dz^{2}
\end{equation}

\section{Some physical parameters}

Using eqn~\eqref{eq:x31} the  directional Hubble's parameter  $H_{x}=H_{y}$,  $H_{z}$ of the model are given by
\begin{equation}\label{eq:x33}
H_{x} = H_{y} = \frac{3c_{1} (c_{1}t+c_{2})^{-1}}{(q+1)(n+2)}
\end{equation}
 and
\begin{equation}\label{eq:x34}
H_{z} = \frac{3nc_{1} (c_{1}t+c_{2})^{-1}}{(q+1)(n+2)}
\end{equation}
equations eqn~\eqref{eq:x33} and eqn~\eqref{eq:x34} give the mean generalized Hubble's parameter H as
\begin{equation}\label{eq:x35}
H = \frac{c (c_{1}t+c_{2})^{-1}}{(q+1)}
\end{equation}
Spatial volume V of the universe is given as
\begin{equation}\label{eq:x36}
V = (c_{1}t+c_{2})^{\frac{3}{(q+1)}}
\end{equation}
The  expansion  scalar $\theta$ of the model is found to be
\begin{equation}\label{eq:x37}
\theta= 3H = \frac{3c_{1}(c_{1}t+c_{2})^{-1}}{(q+1)}
\end{equation}
 The mean anisotropic parameter $A_{m}$ of the model is given as
 \begin{equation}\label{eq:x38}
A_{m} = \frac{2(n-1)^2}{(n+2)^{2}}
 \end{equation}
 The shear scalar $\sigma $ of the model is
 \begin{equation}\label{eq:x39}
 \sigma^{2} = \frac{3c_{1}^2(n-1)^2(c_{1}t+c_{2})^{-2}}{(n+2)^{2}(q+1)^{2}}
 \end{equation}
\subsection{Quark matter for plane symmetric space-time}
Subtracting eqn~\eqref{eq:x24} from eqn~\eqref{eq:x25} we get.
\begin{equation}\label{eq:x40}
-2\frac{\dot{A}\dot{B}}{AB} + \frac{2\ddot{A}}{A} =(1+ 2\lambda)(\rho_{q} +p_{q})
\end{equation}
Applying eqn~\eqref{eq:x31} in eqn~\eqref{eq:x40}, with the help of linear EoS $(p_{q} = \epsilon \rho_{q}, \hspace{.1 cm} 0 \le \epsilon \le 1)$ for $\epsilon = \frac{1}{3}$. We will get pressure for quark matter as follows.
\begin{equation}\label{eq:x41}
p_{q} = \frac{3c_{1}^{2}(c_{1}t+c_{2})^{-2}}{2(1+2\lambda)(q+1)^{2}(n+2)^{2}}[-4n +1 -q(n+2)]
\end{equation}
And energy density for quark matter is given as
\begin{equation}\label{eq:x42}
\rho_{q} = \frac{9c_{1}^{2}(c_{1}t+c_{2})^{-2}}{2(1+2\lambda)(q+1)^{2}(n+2)^{2}}[-4n +1 -q(n+2)]
\end{equation}
Using eqn~\eqref{eq:x31} in eqn~\eqref{eq:x13} we get the Ricci scalar as
\begin{equation}\label{eq:x43}
R= \frac{6c_{1}^{2}(c_{1}t+c_{2})^{-2}}{(q+1)^{2}(n+2)^{2}}[n^{2}(2-q) + n(2-4q)+ (5-4q)]
\end{equation}
\subsection{Strange quark matter for plane symmetric space-time}
Using equation eqn~\eqref{eq:x31} in eqn~\eqref{eq:x40} with the help of EoS parameter given in equation eqn~\eqref{eq:x21}, we will get the pressure and energy density of the strange quark matter as follows
\begin{equation}\label{eq:x44}
p = \frac{3c_{1}^{2}(c_{1}t+c_{2})^{-2}}{2(1+2\lambda)(q+1)^{2}(n+2)^{2}}[-4n +1 -q(n+2)] -B_{c}
\end{equation}
and energy density for quark matter is given as
\begin{equation}\label{eq:x45}
\rho = \frac{9c_{1}^{2}(c_{1}t+c_{2})^{-2}}{2(1+2\lambda)(q+1)^{2}(n+2)^{2}}[-4n +1 -q(n+2)] +B_{c}
\end{equation}

\section{Discussion and conclusion}
In the present paper, we studied the plane symmetric space-time with quark and strange quark matter in the framework of f(R,T) theory of gravity. To decipher the corresponding field equations, we assumed a power law relation between the scale factors. We also considered the special law of variation of Hubble's parameter proposed by \cite{Berman} which yields constant deceleration parameter. The physical behavior of the solutions derived in the previous section is discussed here.
We found that the directional Hubble parameters $H_{x}=H_{y}$, $H_{z}$ as well as mean generalized Hubble parameter (H), expansion scalar ($\theta $), shear scalar ($\sigma $), the spatial volume (V), all these parameters are functions of cosmic time (t). As cosmic time tends to infinity these parameters tends to zero, but when cosmic time is $\frac{-c_{2}}{c_{1}}$ at this time these parameters diverges except the spatial volume V.
\paragraph*{} The spatial volume of this model is zero when cosmic time is $\frac{-c_{2}}{c_{1}}$. Depending on the value of `q', we have following two cases:\\

\begin{figure*}
\includegraphics[scale=0.28]{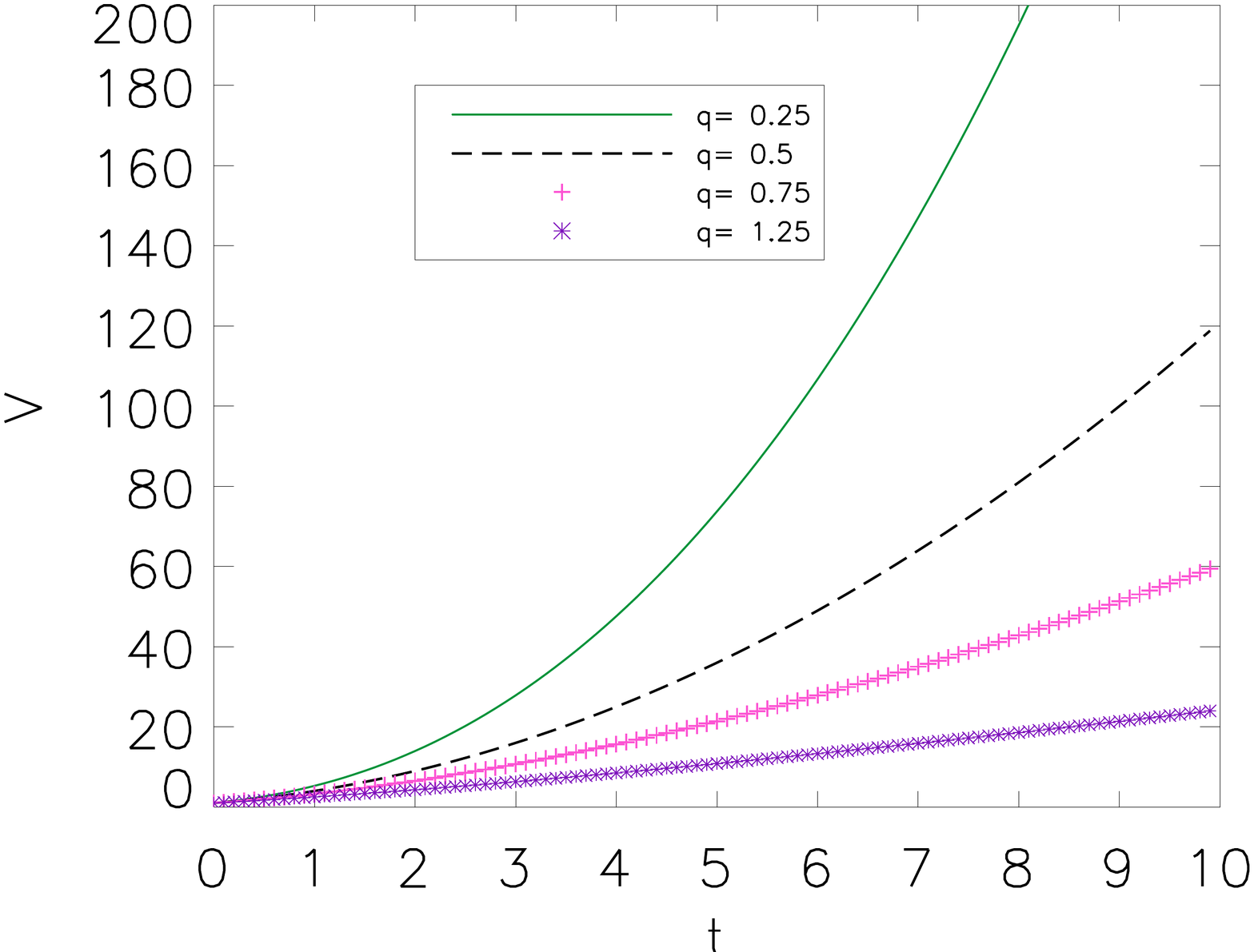}
\includegraphics[scale=0.28]{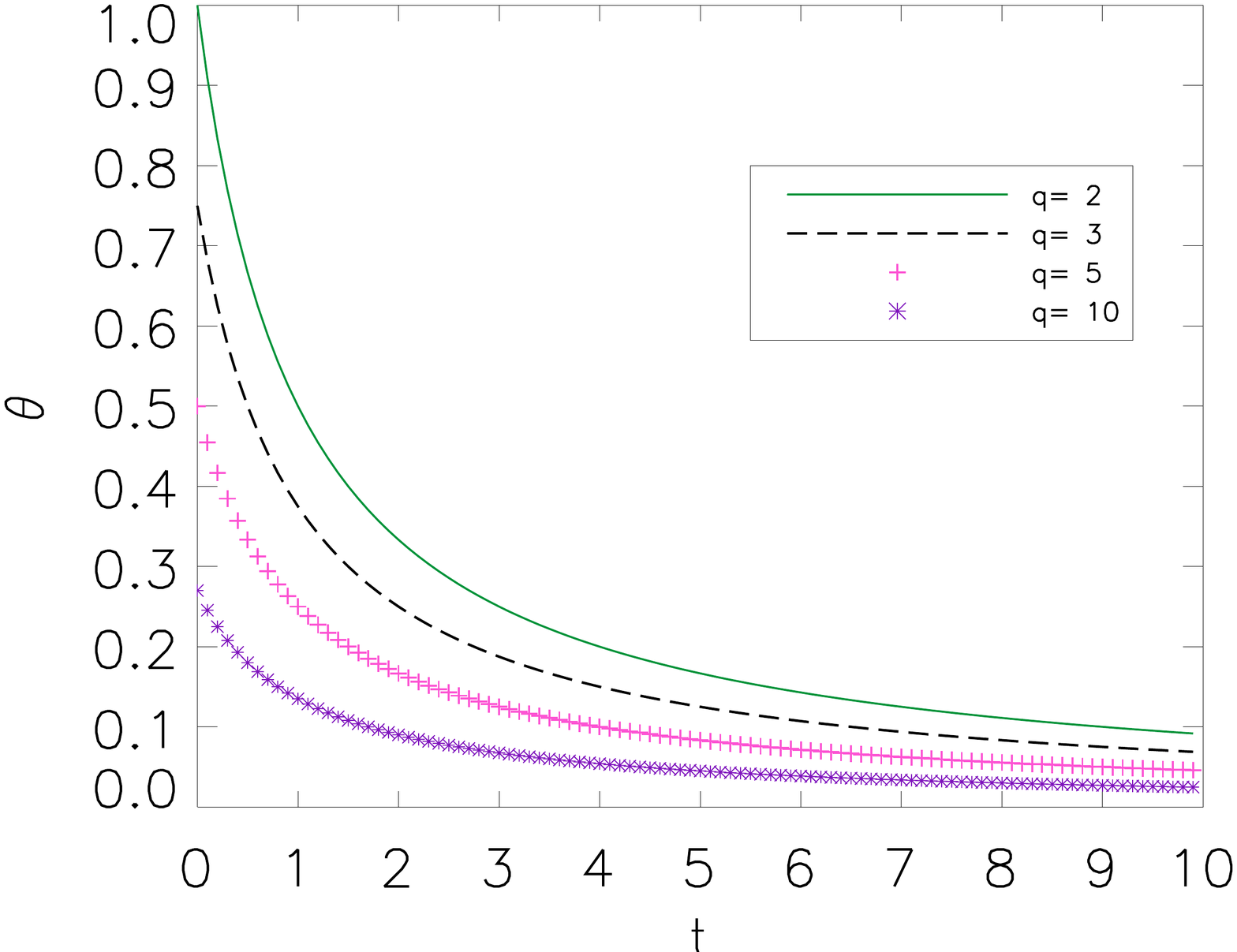}
\caption{Shows the variations of volume (V) (\emph{left}) and scalar expansion $(\theta)$ (\emph{right}) as a function of time (t). V, $\theta$ and t are in arbitrary units. To derive these plots we have used c$_1$ = c$_2$ = 1}\label{fig1}
\vspace{1cm}
\includegraphics[scale=0.28]{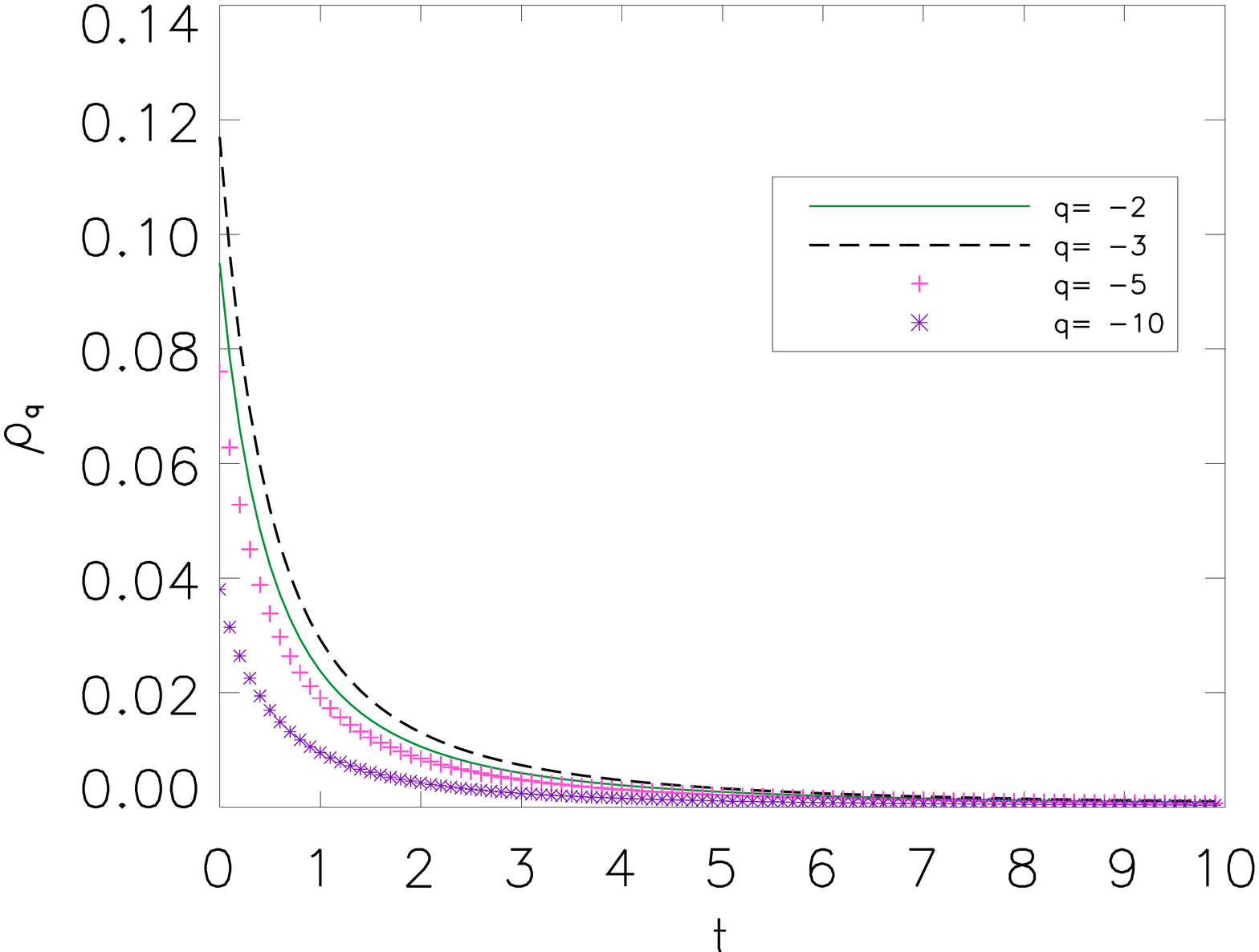}
\includegraphics[scale=0.28]{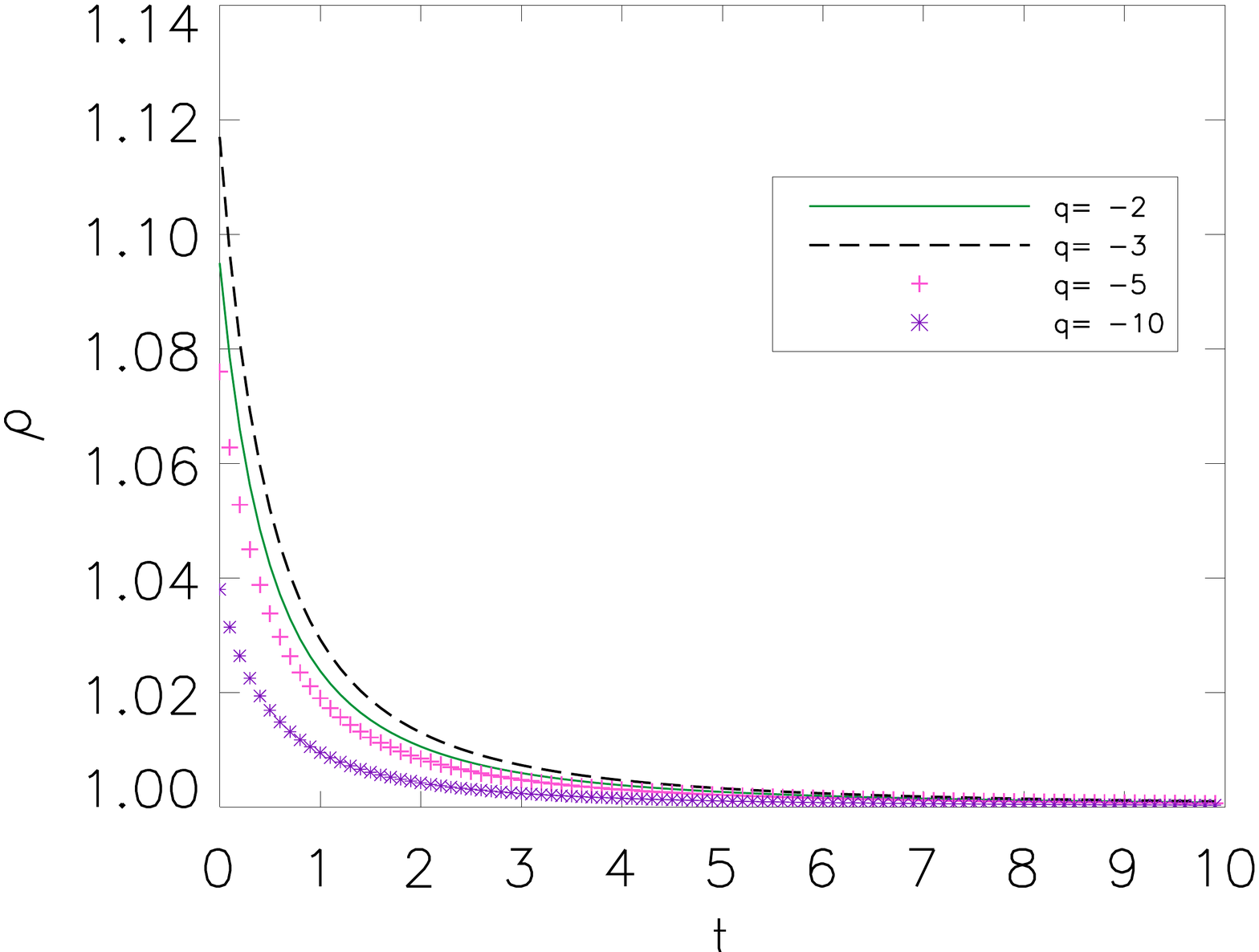}
\caption{Variations of energy density of quark matter (\emph{left}) and of strange quark matter (\emph{right}) as a function of time (t) for $q<-1$ case. All quantities are in arbitrary units. These plots are derived using c$_1$ = c$_2$ = B$_{c}$ = 1 and $n = 2$}\label{fig2}
\vspace{1cm}
\end{figure*}

I) When q $<$ -1: The derived model starts expanding with big bang singularity at $\frac{-c_{2}}{c_{1}}$. 
 The pressure $p_{q}$ and the energy density $\rho_{q}$ for quark matter are finite at $t=0$ and as $t \rightarrow \infty $ both $p_{q}$ and $ \rho_{q} $ becomes zero as shown in left plot of figure~\ref{fig2}. Also, the pressure p and energy density $\rho$ for strange quark matter behaves same as quark matter. The shift in the $\rho$ values than that of $\rho_{q}$ is due to the additional term of Bag constant in the eqn~\eqref{eq:x45}. In this study we chose Bag constant to be unity.\\
 
\begin{figure*}
\includegraphics[scale=0.28]{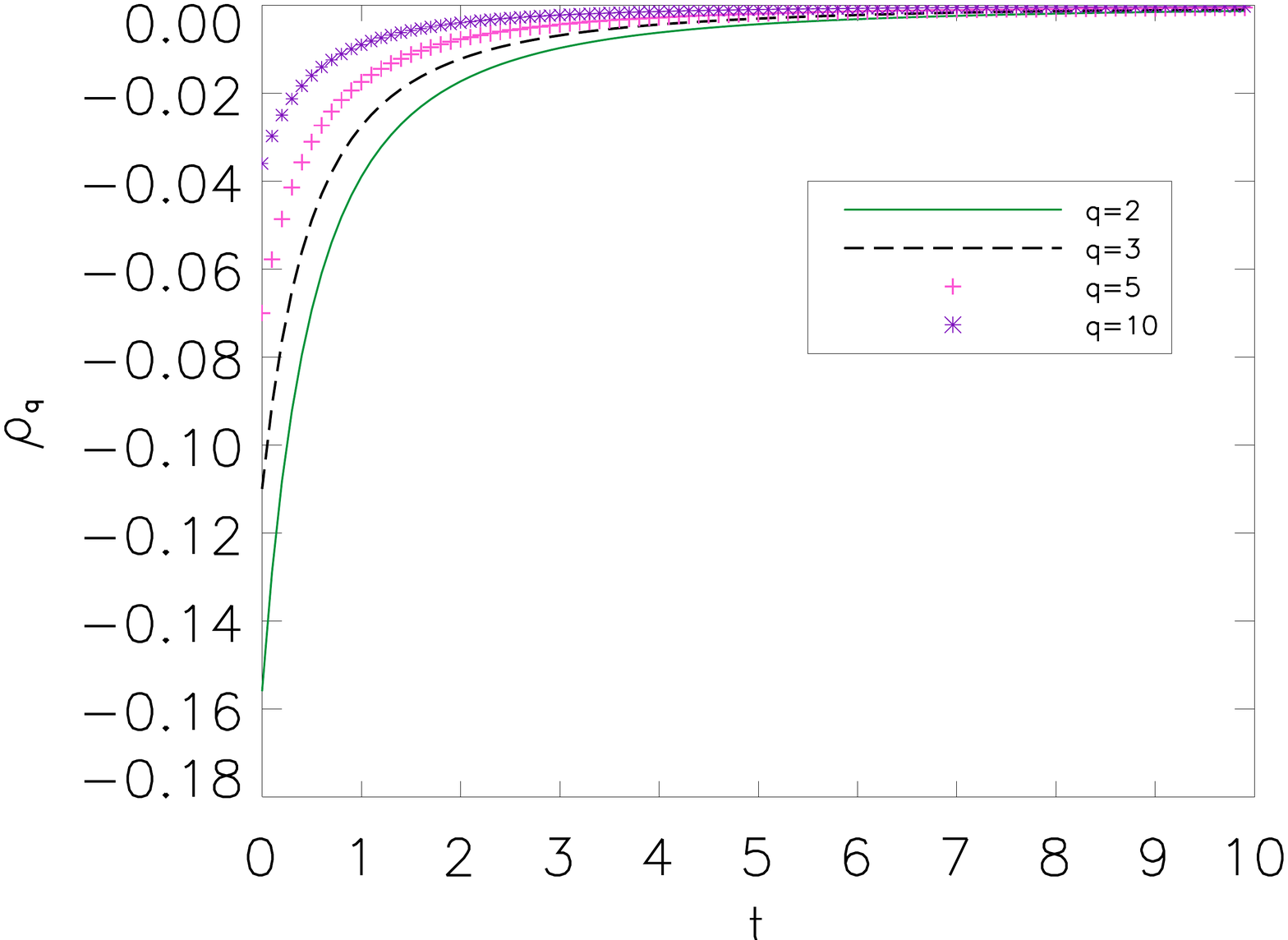}
\includegraphics[scale=0.28]{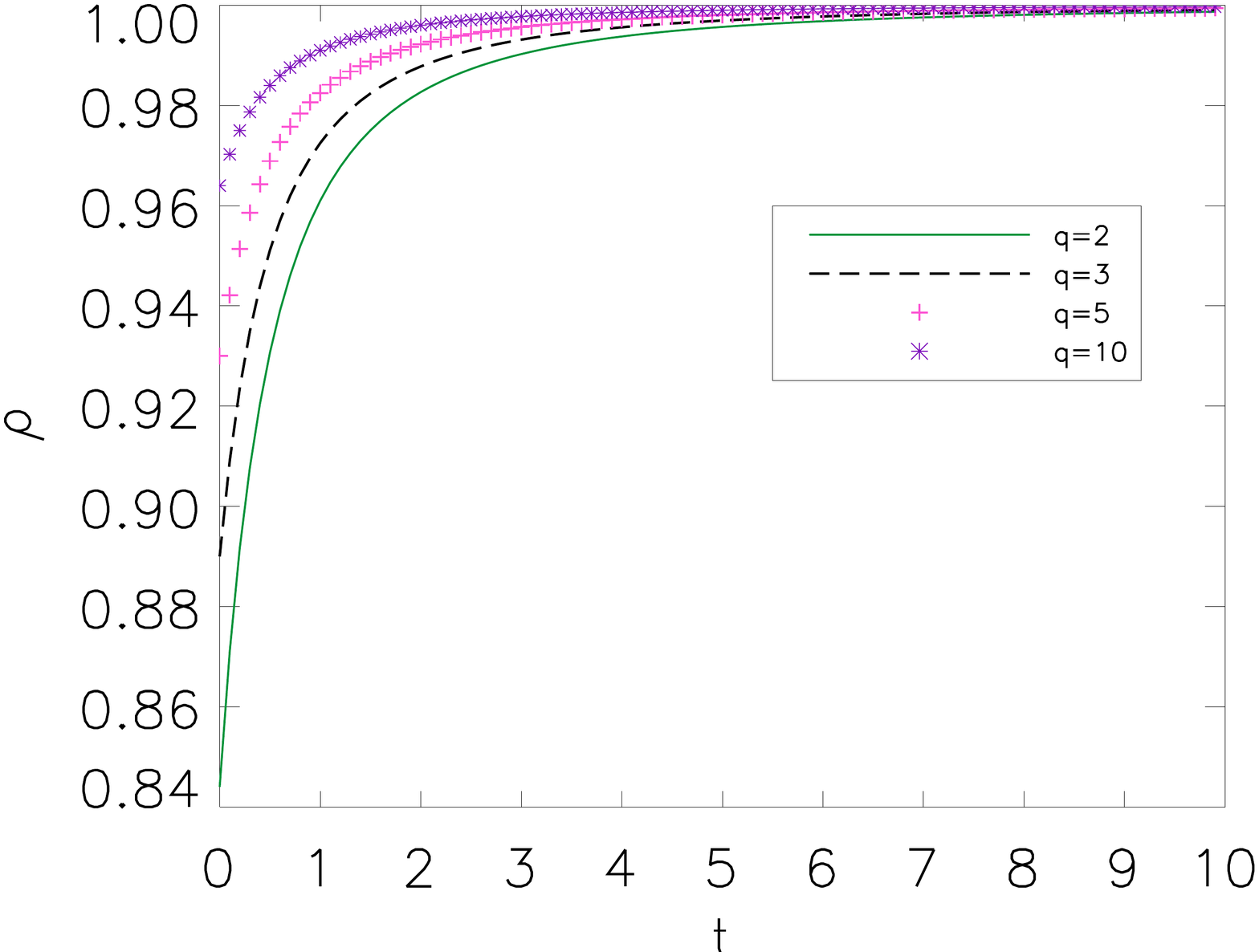}
\caption{Same as figure~\ref{fig2} but for $q>-1$ case. }\label{fig3}
\end{figure*}

II) When q $>$ -1: At cosmic time $t=0$ the model has constant volume and it increases with increase in time and becomes infinite at $t\rightarrow \infty$. As shown in left graph of figure~\ref{fig1}. The pressure $p_{q}$ and energy density $\rho_{q}$ of quark matter  is finite at $t = 0$ and as $t \rightarrow \infty$ they will become zero as shown in left graph of figure ~\ref{fig3}. The behavior of pressure p and energy density $\rho$ for strange quark matter is same as that of quark matter, only it is shifted by Bag constant as shown in right graph of figure~\ref{fig3}.


\paragraph*{} The profiles of energy density and pressure for quark and strange quark matter are same except for the additional Bag constant. For energy density of strange quark we add the Bag constant whereas for pressure we subtract it.
From eqn~\eqref{eq:x38}, the mean anisotropy parameter $A_{m}$ is non-zero for  $ n \neq 1$ and in such case the model does not approach isotropy. But for $n = 1$, the mean anisotropy parameter is zero and the model becomes isotropic. The mean anisotropy parameter ($A_{m}$) is constant throughout the evolution of the universe as it does not depend on the cosmic time. 

\section{Acknowledgements}
We are thankful to the anonymous referee for his/her constructive comments to improve quality of the paper. PKA would like to acknowledge the Department of Science and Technology, New Dehli, India for providing INSPIRE fellowship.

\bibliographystyle{raa}\bibliography{frt_gravity}
\end{document}